\documentclass[11pt]{article}

\usepackage{arxiv}               
\usepackage[hyphens]{url}
\usepackage{graphicx}
\usepackage{multirow}
\usepackage[round]{natbib}
\let\cite\citep                  
\usepackage[labelfont=bf,font=small]{caption}
\usepackage{algorithm}
\usepackage{algorithmic}
\usepackage{newfloat}
\usepackage{amsfonts}
\usepackage{bm}
\usepackage{amsmath}
\usepackage{listings}
\floatstyle{ruled}
\newfloat{listing}{tb}{lst}{}
\floatname{listing}{Listing}

\usepackage{booktabs}
\usepackage{placeins}   

\usepackage[colorlinks=true,linkcolor=BrickRed,citecolor=NavyBlue,urlcolor=NavyBlue]{hyperref}
\usepackage[dvipsnames]{xcolor}

\title{RAG-Audio: Retrieval-Augmented Generation for Faithful Brain-to-Audio Reconstruction}
\shorttitle{RAG-Audio: Faithful Brain-to-Audio Reconstruction}

\author{%
  Ambuj Mehrish \\
  CVML Lab \\
  Ca' Foscari University of Venice \\
  \texttt{ambuj.mehrish@unive.it} \\
  \and
  Sebastiano Vascon \\
  CVML Lab \\
  Ca' Foscari University of Venice \\
  \texttt{sebastiano.vascon@unive.it} \\
}
\date{}

\begin{document}

\maketitle

\begin{abstract}
Brain-to-audio reconstruction is limited by \emph{prior domination}: when a pretrained generator is conditioned on a weak neural signal, it produces realistic but stimulus-inaccurate audio. We introduce RAG-Audio, which decodes fMRI into a semantic audio embedding, retrieves a matching real-audio exemplar, and initializes the frozen generator's sampling trajectory from that exemplar while retaining the decoded embedding as conditioning. On Brain2Music, RAG-Audio improves 10-way stimulus identification from $0.14$--$0.18$ for direct generation, near the $0.10$ chance level, to $0.40$--$0.43$, comparable to retrieval. It also reduces Fr\'echet Audio Distance by roughly an order of magnitude, from $13.49$ to $1.25$ for AudioLDM. RAG-Audio approaches nearest-neighbor retrieval in identification while remaining generative; its higher FAD is expected because retrieval directly replays real audio. An autoregressive negative control, which lacks an initializable latent trajectory, shows no comparable gain, attributing the improvement to trajectory initialization. These results suggest that retrieval-guided initialization can mitigate prior domination in brain-to-audio generation.
\end{abstract}

\section{Introduction}
Reconstructing perceptual experience from brain activity is a demanding challenge for modern generative models. In vision, recent methods can recover recognizable images from functional magnetic resonance imaging (fMRI), even though fMRI measures stimulus-related changes in blood oxygenation rather than neural activity directly~\cite{nishimoto2011reconstructing,takagi2023high,scotti2023reconstructing,huo2024neuropictor}. Audio reconstruction has received far less attention, especially for music. Most brain-to-audio systems use a two-stage pipeline: they first map the fMRI response to a semantic audio embedding and then use this embedding to guide a pretrained audio generator~\cite{denk2023brain2music,ciferri2025reconstructing,park2023sound}. Brain2Music \cite{denk2023brain2music}, for example, predicts a joint music–language embedding \cite{huang2022mulan} from fMRI and supplies it to a text-to-audio model. This design appears reasonable because it separates neural decoding from waveform generation. Yet it also assumes that the generator will preserve whatever stimulus information survives the decoding stage.
\begin{figure}[!htb]
    \centering
    \includegraphics[width=\linewidth]{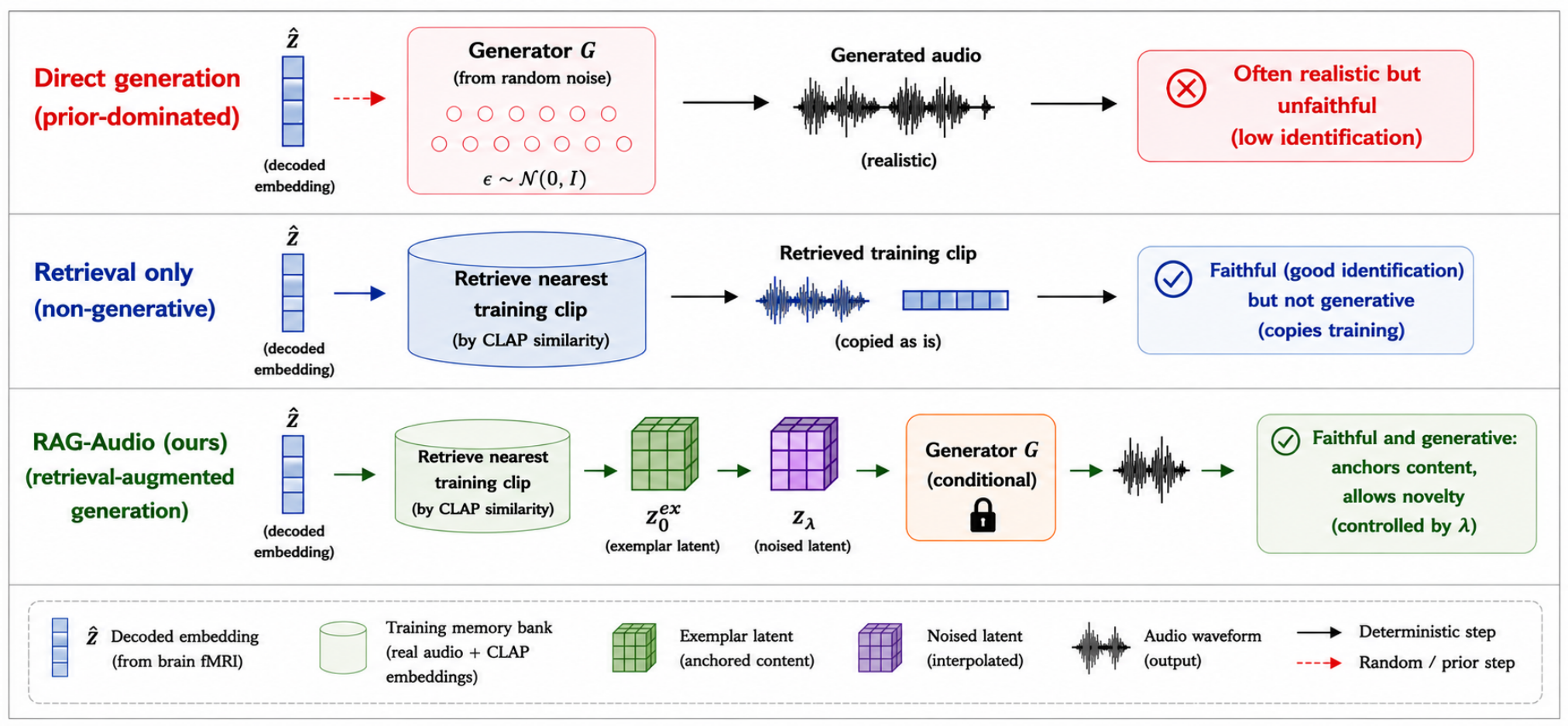}
    \caption{\textbf{Overview of RAG-Audio and two baselines.} Given a brain-decoded CLAP embedding $\hat{z}$, direct generation samples from noise and may be dominated by the generator prior; retrieval returns the nearest training clip but is non-generative. RAG-Audio initializes the frozen generator from the exemplar latent $z_0$ perturbed to an intermediate time, $z_{t_0}$, preserving stimulus-related structure while the anchoring strength $s$ (with $t_0=sT$) controls the trade-off between faithfulness and novelty.}
    \label{fig:overview}
\end{figure}

Our results suggest that this assumption often fails. We refer to this failure as \emph{prior domination}: when a powerful pretrained generator receives a weak or noisy brain-derived condition, its learned distribution over plausible audio can outweigh the evidence carried by that condition. This is the conditioning-strength problem familiar from diffusion guidance, where a guidance weight balances a conditional signal against the unconditional prior~\cite{dhariwal2021diffusion,ho2022classifier}; here the brain-derived condition is fixed and weak, so no reweighting of a text prompt can recover it. The generated audio may sound realistic but still fail to match the music the participant actually heard. This gap is clear on the public Brain2Music dataset~\cite{denk2023brain2music}, which contains fMRI recordings from five subjects listening to GTZAN music clips~\cite{tzanetakis2002musical}. In our re-implementation, decoding fMRI into a CLAP embedding identifies the heard clip with 0.43 accuracy in a 10-way test, well above the $0.10$ chance level. However, when the same embedding is passed to AudioLDM~\cite{liu2023audioldm}, TangoFlux~\cite{hung2024tangoflux}, or MusicGen~\cite{copet2023simple}, identification accuracy for the generated audio falls to $0.14$--$0.18$, including in the published Brain2Music generation setting~\cite{denk2023brain2music}. The conditioning signal therefore contains recoverable stimulus information, but direct generation retains little of it. Brain-to-image studies have reported a related qualitative pattern, with realistic reconstructions that drift from the target~\cite{takagi2023high,scotti2024mindeye2} (Fig.~\ref{fig:recon}), although it remains unclear whether the same behavior can be measured systematically in audio or whether it depends on a particular generator.

We investigate a targeted way to reduce prior domination (Figure~\ref{fig:overview}). Instead of asking the generator to reconstruct the stimulus from the decoded embedding alone, we first retrieve the training audio clip whose CLAP embedding is closest to the decoded representation. We then use this clip to initialize the generator at an intermediate point in its sampling trajectory. For a diffusion-based generator such as AudioLDM~\cite{liu2023audioldm}, we follow the SDEdit principle~\cite{mengsdedit} by adding partial noise to the exemplar latent before reverse diffusion. For a flow-based generator such as TangoFlux~\cite{hung2024tangoflux}, we use the corresponding deterministic interpolation along the rectified-flow trajectory~\cite{liu2023flow,lipman2022flow,rout2024semantic,kulikov2025flowedit}. The generator can therefore refine the retrieved audio rather than synthesize entirely from its prior. The anchoring strength controls how closely the output follows the exemplar: weaker perturbation preserves more of it, while stronger perturbation allows greater variation and novelty. This approach differs from plain retrieval because the final output does not simply copy the stored clip. It also differs from direct brain-conditioned generation because sampling begins from stimulus-related audio structure. We therefore view exemplar anchoring as a tunable trade-off between faithfulness and novelty, rather than as a replacement for retrieval or unconstrained generation~\cite{lewis2020retrieval,blattmann2022semi,yuan2024retrieval}.

These design choices yield a consistent improvement: anchoring recovers identification to the level of a strong non-generative retrieval baseline~\cite{ciferri2025reconstructing} while keeping the output newly generated, and an autoregressive negative control (MusicGen~\cite{copet2023simple}), which exposes no initializable trajectory, shows no comparable gain isolating trajectory initialization rather than retrieval as the mechanism. Section~\ref{sec:results} reports the full results. In summary, our contributions are:
\begin{itemize}
\item We identify and quantify \emph{prior domination}, a failure mode the decode-then-generate recipe inherits and that sharpens as pretrained generators grow stronger.
\item We introduce exemplar anchoring, a retrieval-augmented scheme that initializes the generator partway along its sampling trajectory from a retrieved real-audio exemplar, recovering retrieval-level identification while keeping generation genuinely generative and reducing FAD by $\sim10.8 \times$ for AudioLDM and ${\sim}3.3\times$ for TangoFlux.
\item We use the autoregressive generator MusicGen~\cite{copet2023simple} as a negative control, attributing the recovery to latent-trajectory initialization rather than to retrieval alone.
\item We provide an anatomical validity check that localizes the decoder signal to auditory cortex.
\end{itemize}

\section{Related Work}
\paragraph{Brain-to-audio reconstruction.}
Perceived sound has been decoded from ECoG, MEG, EEG, and fMRI, spanning speech, music, and semantic language~\cite{anumanchipalli2019speech,metzger2023high,bellier2023music,defossez2023decoding,postolache2025naturalistic,tang2023semantic}. For fMRI music, most methods decode an audio representation and condition a pretrained generator~\cite{denk2023brain2music,park2023sound,liu2024reverse}. Existing systems thus range from nearest-neighbour retrieval~\cite{ferrante2024r} to unconstrained generation, including prior-guided diffusion~\cite{ciferri2025reconstructing}. We re-implement both endpoints in a common evaluation framework and bridge them through exemplar anchoring.

\paragraph{Prior domination and cortical localization.}
Brain-to-image reconstruction exhibits a closely related tension. Latent-diffusion decoders reconstruct images from fMRI on the Natural Scenes
Dataset~\cite{allen2022massive} by mapping activity into a generator's conditioning space~\cite{takagi2023high, chen2023seeing}, and the MindEye family separates a retrieval pathway from a diffusion-prior reconstruction pathway~\cite{scotti2023reconstructing, scotti2024mindeye2}; a recurring observation is that outputs look realistic yet drift from the stimulus when the brain signal is weak relative to the generator prior. We term this regime prior domination and, unlike the largely qualitative image-domain reports, quantify it for audio and show it holds across generator families. Our decoder-localization analysis follows voxelwise modeling of auditory cortex, where deep-network features predict responses along the auditory hierarchy and music-selective populations occupy superior temporal cortex~\cite{kell2018task, norman2015distinct, tuckute2023many}.

\paragraph{Audio generators and embeddings.}
Brain-to-image reconstruction faces a similar trade-off. Diffusion-based fMRI decoders can produce realistic images but often drift from the stimulus when the generator prior overwhelms weak neural evidence~\cite{allen2022massive,takagi2023high,chen2023seeing,scotti2023reconstructing,scotti2024mindeye2}. We quantify this \emph{prior domination} for audio across generator families and localize decoder contributions using established models of auditory-cortex organization~\cite{kell2018task,norman2015distinct,tuckute2023many}.

\paragraph{Editing and retrieval-augmented generation.}
SDEdit starts from a guide signal, adds noise up to a chosen intermediate diffusion step, and then denoises from that point to produce a new sample~\cite{mengsdedit}. The noise level controls the balance between preserving the guide and allowing the generator to introduce new content~\cite{ho2020denoising,song2020score,dhariwal2021diffusion,ho2022classifier,rombach2022high}. Related methods guide frozen diffusion models with reference images~\cite{choi2021ilvr,lugmayr2022repaint}, while retrieval-augmented generators condition on external examples across language, image, and audio synthesis~\cite{khandelwal2019generalization,lewis2020retrieval,borgeaud2022improving,blattmann2022semi,yuan2024retrieval}. RAG-Audio combines these ideas by initializing generation from a retrieved audio exemplar, preserving stimulus-relevant content without copying the retrieved clip verbatim.
\begin{figure*}[!htb]
    \centering
    \includegraphics[width=\linewidth]{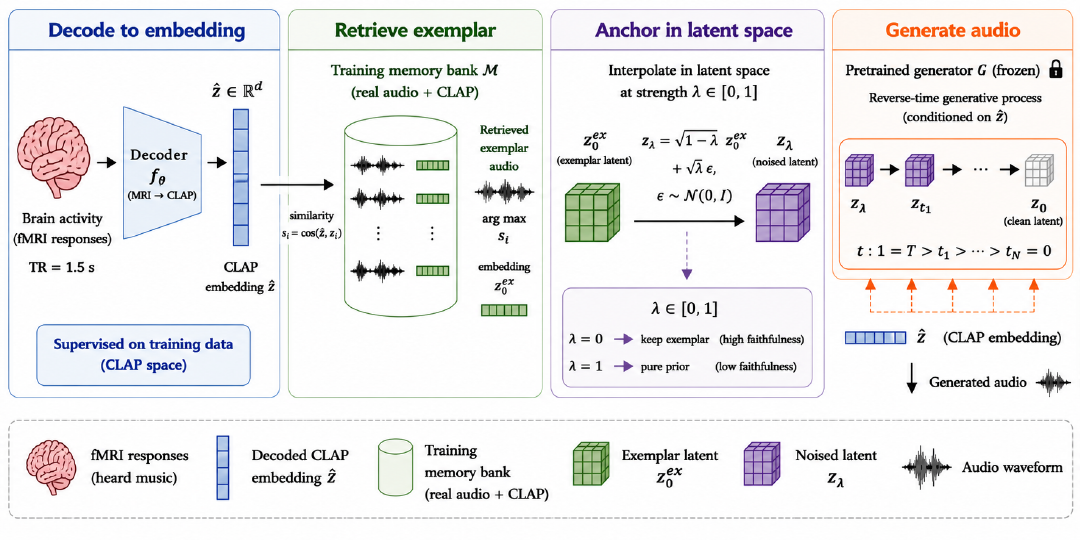}
    \caption{\textbf{The RAG-Audio pipeline.} The decoder $f_\theta$ maps an fMRI response $x$ to a CLAP embedding $\hat{z}$; the nearest exemplar $a^{\ast}$ is retrieved and encoded to $z_0$, perturbed to an intermediate time $t_0=sT$ ($z_{t_0}$), and the frozen generator denoises from $z_{t_0}$ conditioned on $\hat{z}$ to produce $\hat{a}$. The strength $s$ trades faithfulness against novelty.}
\label{fig:method}
\end{figure*}
\section{Problem Setup}
Let $a$ denote an audio clip heard by a subject while functional magnetic resonance imaging (fMRI) records the corresponding brain response $x$. Given a paired dataset $\mathcal{D}={(x_i,a_i)}_{i=1}^{M}$, the goal is to reconstruct the stimulus associated with a held-out scan. Since fMRI is an indirect and temporally coarse measurement of neural activity, we formulate reconstruction in a semantic audio space rather than at the waveform level. Each stimulus is represented using CLAP~\cite{wu2023large}:

\begin{equation}
z_a=\mathrm{CLAP}(a)\in\mathbb{R}^{512}.
\end{equation}

Following the standard decode-then-generate pipeline \cite{denk2023brain2music}, a learned decoder $f$ maps the fMRI response to a predicted semantic embedding:

\begin{equation}
\hat{z}=f(x).
\end{equation}

A frozen pretrained audio generator $G$ then synthesizes the reconstruction:

\begin{equation}
\hat{a}=G(\hat{z}).
\end{equation}

We defer the architectures and training procedures for $f$ and $G$ to the Method section.

We measure semantic faithfulness using $N$-way identification. Let $a^{+}$ be the true stimulus, and let $\mathcal{A}^{-}={a_1^{-},\ldots,a_{N-1}^{-}}$ contain $N-1$ distractors. Writing $s(\cdot,\cdot)$ for cosine similarity, the identification accuracy is

\begin{equation}
\begin{split}
\mathrm{acc}_{N} = \mathbb{E}\Bigg[ \mathbb{I}\Bigg(
& s\!\left(\mathrm{CLAP}(\hat{a}), \mathrm{CLAP}(a^{+}) \right) \\ & > \max_{a^{-}\in\mathcal{A}^{-}} s\!\left( \mathrm{CLAP}(\hat{a}), \mathrm{CLAP}(a^{-}) \right) \Bigg) \Bigg].
\end{split}
\label{eq:nway-identification}
\end{equation}

Chance performance is $1/N$. This metric evaluates whether the reconstruction preserves the semantic content of the heard stimulus; it does not measure sample-accurate waveform recovery.

The two stages can contribute unequally to the final output. The generator $G$ represents a strong prior $p(a)$ learned from large audio corpora, whereas the brain-derived condition $\hat{z}$ is comparatively weak and noisy. We use \emph{prior domination} to describe the regime in which $G(\hat{z})$ remains acoustically plausible but depends only weakly on $\hat{z}$, behaving approximately like an unconditional sample from the generator prior. Generation may therefore discard stimulus information retained by the decoded embedding. Our objective is to preserve the realism supplied by $G$ while improving the semantic faithfulness of its output.
\section{Method}

We propose \emph{RAG-Audio}, a retrieval-augmented pipeline for brain-to-audio reconstruction (Fig.~\ref{fig:method}). RAG-Audio first decodes an fMRI response into a CLAP embedding, retrieves the nearest real training exemplar from a memory bank, and then initializes a frozen generator's sampling trajectory from that exemplar at an intermediate time. For AudioLDM, this follows the SDEdit principle of partially noising the exemplar latent before reverse diffusion \cite{mengsdedit}, while TangoFlux~\cite{hung2024tangoflux} applies the corresponding deterministic interpolation along its rectified-flow trajectory. The retrieved clip supplies plausible audio structure that can survive the generator's prior, turning prior domination into a controllable trade-off between faithfulness to the retrieved content and generative novelty.

\subsection{fMRI-to-CLAP Decoder}

For each stimulus audio clip $a$, we use its CLAP representation \cite{wu2023large} as the decoding target, with $z_a=\mathrm{CLAP}(a)\in\mathbb{R}^{512}$. We apply rigid motion correction with ANTsPy~\cite{avants2011reproducible} and retain each subject's data in native space. To account for the haemodynamic response, we use a lag of $l=3$ repetition times (TRs), corresponding to $4.5$ seconds, and a temporal window of $w=8$ TRs.

We rank voxels according to their association with the CLAP target. For voxel $v$, let $x_v$ denote its temporal response and let $z^{(d)}$ denote the $d$-th CLAP dimension. We define the relevance score as

\begin{equation}
s_v=\max_{d\in\{1,\ldots,512\}}\big|\mathrm{corr}(x_v,z^{(d)})\big|. \label{eq:vox}
\end{equation}

We retain the $K=2000$ highest-scoring voxels. The same scores define the localization map analyzed in Figure~\ref{fig:atlas}.

The decoder $f_\theta:\mathbb{R}^{2000}\rightarrow\mathbb{R}^{512}$ maps the selected fMRI features to a predicted embedding $\hat z=f_\theta(x)$. Rather than minimizing dimension-wise reconstruction error, we train $f_\theta$ with InfoNCE~\cite{oord2018representation} so that decoded scans remain close to their paired audio embeddings and separated from other clips in the batch. For a batch of size $B$ and temperature $\tau=0.07$, the objective is

\begin{equation}
\mathcal{L}=-\frac{1}{B}\sum_{i=1}^{B}\log
\frac{\exp\!\big(\cos(f_\theta(x_i),z_{a_i})/\tau\big)}
{\sum_{j=1}^{B}\exp\!\big(\cos(f_\theta(x_i),z_{a_j})/\tau\big)}. \label{eq:infonce}
\end{equation}

This objective directly optimizes the retrieval geometry used for semantic identification. We compare it with ridge, linear, and MLP decoders in Figure~\ref{fig:decoder}; the contrastive decoder is selected because its training criterion matches the downstream identification objective.

\subsection{Memory Bank and Retrieval}

We construct a memory bank $\mathcal{B}=\{z_a:a\in\mathcal{A}_{\mathrm{train}}\}$ from CLAP embeddings of real training-set audio, where $\mathcal{A}_{\mathrm{train}}$ denotes the training clips. Test stimuli are excluded. Each bank entry retains a reference to its corresponding audio clip.

Given a held-out fMRI response $x$, the decoder produces $\hat z=f_\theta(x)$. We retrieve the training exemplar whose CLAP embedding has the largest cosine similarity to $\hat z$:

\begin{equation}
a^\ast=\arg\max_{a\in\mathcal{B}}\ \cos(\hat z,z_a). \label{eq:retr}
\end{equation}

Here, $a\in\mathcal{B}$ indexes the audio clip associated with a bank embedding. Returning $a^\ast$ directly defines the \textsc{Retrieval} baseline, corresponding to the published linear-contrastive fMRI-to-CLAP retrieval setting \cite{ferrante2024r}. This baseline provides a strong non-generative reference: it can preserve stimulus-level content, but its output is a verbatim training clip. RAG-Audio instead treats $a^\ast$ as an anchor from which to generate a new reconstruction.

\subsection{Exemplar-Anchored Generation via Intermediate-Time Initialization}

Let $E$ denote the encoder of a frozen latent generative model. We first map the retrieved exemplar to the generator latent, $z_0=E(a^\ast)$. A anchoring strength $s\in(0,1]$ determines the intermediate initialization time $t_0=sT$, where $T$ is the full denoising horizon.

For a latent-diffusion generator such as AudioLDM
\cite{liu2023audioldm}, the anchoring strength
\(s \in (0,1]\) determines the diffusion initialization step
\begin{equation}
t_0 = \left\lfloor sT \right\rfloor,
\label{eq:diffusion_time}
\end{equation}
where \(T\) denotes the total number of diffusion steps. Following the SDEdit principle \cite{mengsdedit}, we partially corrupt the exemplar latent \(z_0\) according to the forward diffusion process and then run the reverse process from \(t_0\), conditioned on the decoded embedding
\(\hat{z}\):
\begin{equation}
z_{t_0} = \sqrt{\bar{\alpha}_{t_0}}\,z_0 + \sqrt{1-\bar{\alpha}_{t_0}}\,\epsilon, \qquad \epsilon \sim \mathcal{N}(0,I).
\label{eq:sdedit}
\end{equation}

Flow based generator such as TangoFlux~\cite{hung2024tangoflux} follows a normalized rectified-flow trajectory. For this model, the anchoring strength directly determines the initialization time:
\begin{equation}
\tau_0 = s.
\label{eq:flow_time}
\end{equation}
We initialize the trajectory by deterministically interpolating between the exemplar latent and Gaussian noise~\cite{song2020denoising}:
\begin{equation}
z_{\tau_0}
=
(1-\tau_0)z_0
+
\tau_0\epsilon,
\qquad
\epsilon \sim \mathcal{N}(0,I),
\label{eq:flow_init}
\end{equation}
and then integrate the learned flow from \(\tau_0\) to \(0\).

\begin{figure}
\centering
  \includegraphics[width=0.75\linewidth]{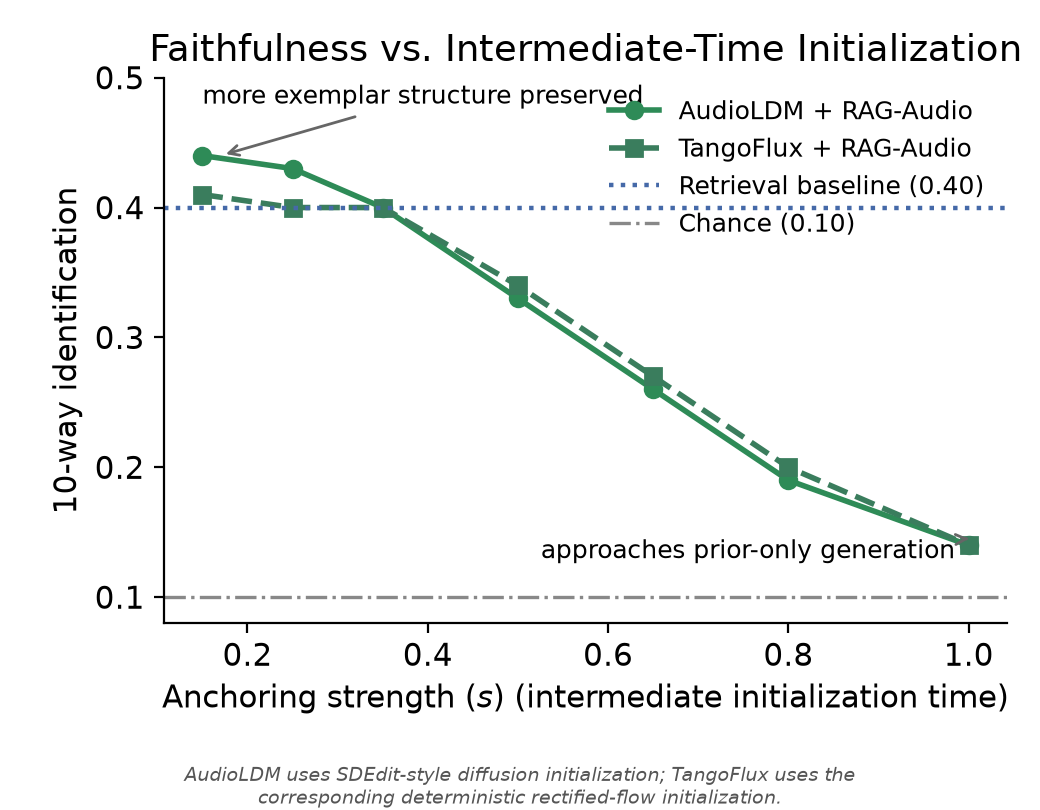}
\caption{Ten-way identification versus anchoring strength $s$: low $s$ stays near the retrieval bound, high $s$ decays to the $0.10$ chance level; working range $s\!\approx\!0.25$--$0.40$.}
  \label{fig:curve}
\end{figure}

The strength $s$ controls how much exemplar structure remains at the start of generation. As $s\rightarrow 0$, the output approaches $a^\ast$, favoring faithfulness at the cost of novelty. As $s\rightarrow 1$, the initialization increasingly erases the exemplar and approaches direct generation from the model prior. Sweeping $s$ therefore traces the faithfulness--novelty continuum shown in Figure~\ref{fig:curve}; empirically, we consider the working range $s\approx0.25$--$0.40$.

This mechanism requires a continuous latent trajectory that can be initialized through Equation~\eqref{eq:sdedit}. Autoregressive MusicGen \cite{copet2023simple} does not provide such a latent. Its retrieved exemplar can enter only through melody conditioning, which remains subject to the autoregressive token prior. We use this case as a negative control in the Results section to distinguish the effect of intermediate-time trajectory initialization from the mere availability of an exemplar.
\begin{table}[!htb]
\centering
\small
\caption{Main comparison on Brain2Music ($n{=}300$; $10$-way chance $0.10$), all methods run in a single harness. Exemplar anchoring lifts identification from $0.14$--$0.18$ (direct) to the retrieval level ($0.40$--$0.43$) and cuts FAD by up to ${\sim}10\times$ for AudioLDM and TangoFlux, but not for autoregressive MusicGen.}
\label{tab:main}
\begin{tabular}{llccc}
\toprule
Method & Type & Gen & 10-way$\uparrow$ & FAD$\downarrow$\\
\midrule
\multicolumn{5}{l}{\emph{Published baselines (re-implemented)}}\\
R\&B~\cite{ferrante2024r}      & retrieval   & no  & 0.40 & \textbf{0.89}\\
MusicGen~\cite{copet2023simple}(decode$\to$gen) & direct & yes & 0.18 & 8.06\\
\multicolumn{5}{l}{\emph{Direct generation, prior-only (ours)}}\\
AudioLDM~\cite{liu2023audioldm}                      & direct     & yes & 0.14 & 13.5\\
TangoFlux~\cite{hung2024tangoflux}                     & direct     & yes & 0.14 & 7.89\\
MusicGen~\cite{copet2023simple}                     & direct     & yes & 0.18 & 8.06\\
\multicolumn{5}{l}{\emph{RAG-Audio (ours)}}\\
AudioLDM + RAG                & retr.-aug. & yes & \textbf{0.43} & 1.25\\
TangoFlux + RAG               & retr.-aug. & yes & 0.40 & 2.36\\
MusicGen + RAG                & retr.-aug. & yes & 0.20 & 6.21\\
\bottomrule
\end{tabular}
\end{table}
\section{Experimental Setup}
\paragraph{Dataset.}
We evaluate on Brain2Music, released as OpenNeuro ds003720 \cite{nakai2022music,denk2023brain2music}. The dataset contains fMRI recordings from five subjects listening to 15-second music clips drawn from the ten GTZAN genres~\cite{tzanetakis2002musical}, with a repetition time of $\mathrm{TR}=1.5,\mathrm{s}$. We follow the official subject-wise split of 480 training clips and 60 held-out test clips. All metrics are averaged across subjects, corresponding to a pooled set of $n=300$ test generations.

\paragraph{Audio generators.}
We consider three pretrained generators: AudioLDM (\texttt{cvssp/audioldm-s-full-v2}) \cite{liu2023audioldm}, TangoFlux (\texttt{declare-lab/TangoFlux}) \cite{hung2024tangoflux}, and MusicGen (\texttt{facebook/musicgen-melody}) \cite{copet2023simple}. All generator parameters remain frozen. AudioLDM and TangoFlux use latent-diffusion and rectified-flow formulations, respectively, and therefore support intermediate-time trajectory initialization. MusicGen is autoregressive and serves as a mechanism control for which exemplar initialization is unavailable.

\paragraph{Comparison arms.}
We compare three reconstruction strategies. \emph{Direct} performs brain-conditioned generation using only the pretrained generative prior. \emph{Retrieval} returns the nearest real clip from the training bank~\cite{ferrante2024r}. \emph{RAG-Audio} applies our intermediate-time exemplar-anchoring procedure. All methods, including published baselines, are run within a single evaluation harness using the same test scans, candidate sets, checkpoints, and metric implementations. Their reported values are therefore directly comparable.

\paragraph{Faithfulness.}
Semantic faithfulness is measured by the $N$-way identification of Eq.~\ref{eq:nway-identification}, computed in CLAP space using the LAION checkpoint \texttt{laion/clap-htsat-unfused}. We report $N\in\{2,5,10,50\}$, with corresponding chance levels of $0.50$, $0.20$, $0.10$, and $0.02$.

\paragraph{Realism and novelty.}
Audio realism is evaluated using FAD computed with VGGish features~\cite{kilgour2019frechet,hershey2017cnn}, together with FD and KL divergence computed from the PANNs Cnn14 audio tagger~\cite{kong2020panns}. Lower values indicate better agreement with real audio. We quantify novelty as one minus the cosine similarity between each generated clip and its nearest clip in the training bank. Plain retrieval has novelty near zero by construction because it returns a training clip verbatim. We additionally report pairwise diversity among generated samples.
\begin{figure}
\centering
  \includegraphics[width=0.7\linewidth]{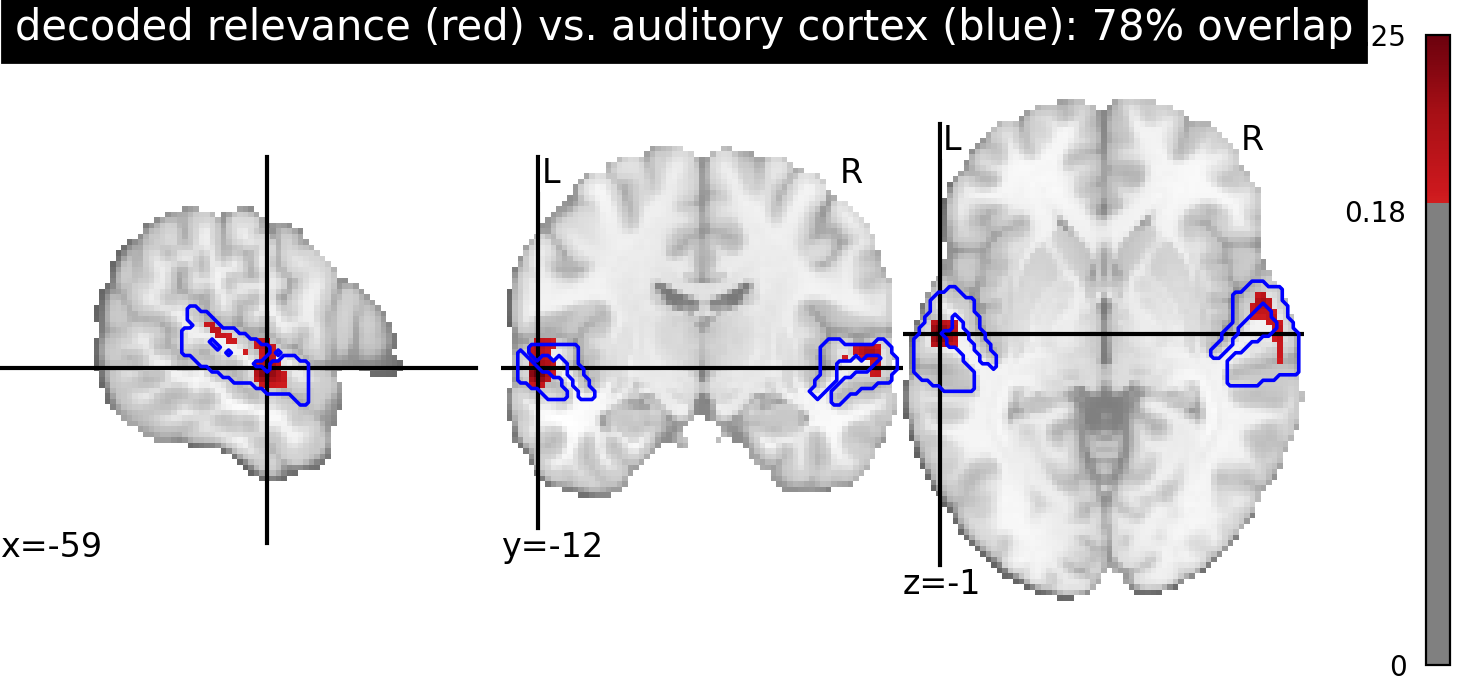}
 \caption{Decoder localization: voxel relevance scores (red) overlap the Harvard--Oxford auditory-cortex mask (blue), with $78\%$ of top voxels in bilateral superior temporal gyrus.}
  \label{fig:atlas}
\end{figure}

\paragraph{Implementation.} We apply rigid-motion correction with ANTsPy~\cite{avants2011reproducible}, select the top $2000$ voxels, sweep the anchoring strength over $\{0.2,\ldots,0.5\}$, and train the contrastive decoder with temperature $\tau=0.07$. Complete preprocessing, optimization, and generation settings are provided in Appendix~\ref{app:impl}.
\begin{figure}\centering
  \includegraphics[width=0.7\linewidth]{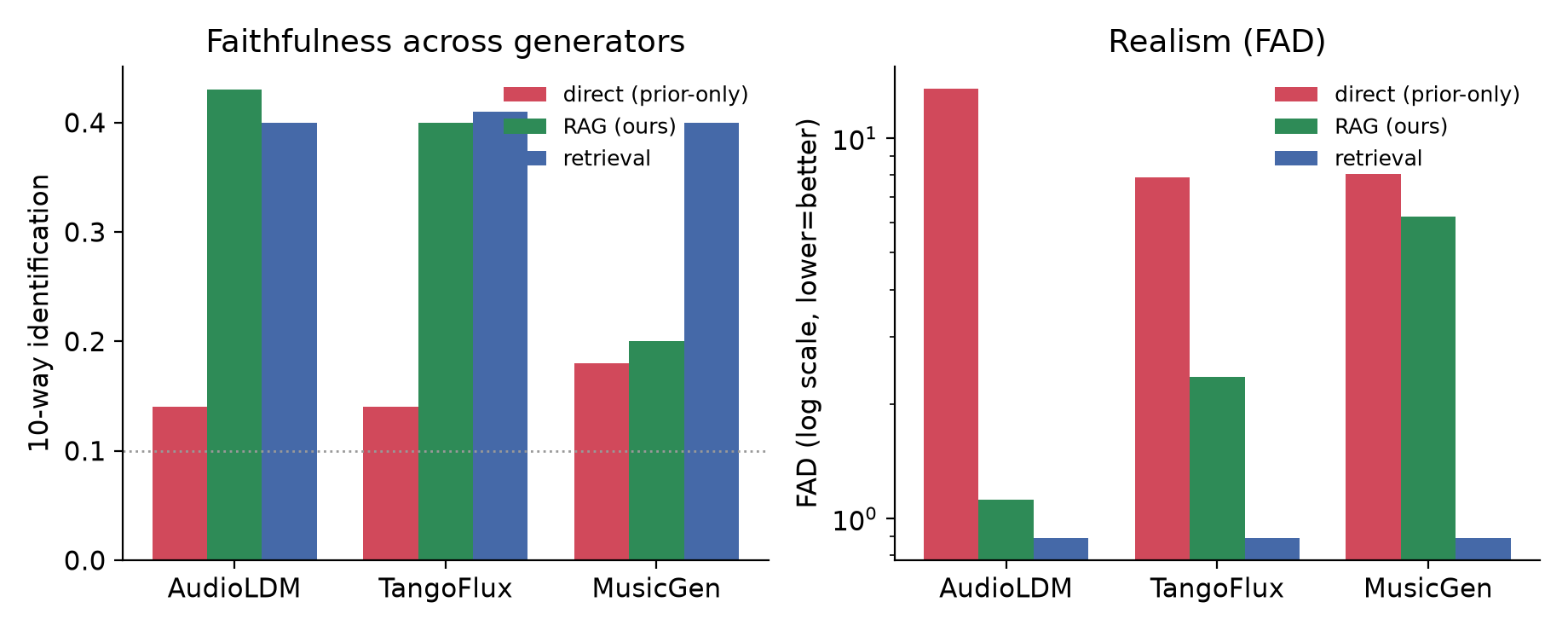}
\caption{Ten-way identification and FAD per generator: anchoring restores AudioLDM and TangoFlux to retrieval level but not autoregressive MusicGen.}
  \label{fig:cross}
\end{figure}
\section{Results}
\label{sec:results}
\paragraph{Decoder Performance} The contrastive fMRI-to-CLAP decoder achieves the strongest stimulus identification despite the lowest dimension-wise embedding correlation. Averaged across five subjects, it reaches $0.43$ accuracy in 10-way identification (chance $0.10$) and $0.80$ in 2-way identification, with every subject above 10-way chance. It outperforms ridge, linear, and MLP decoders even though its Pearson correlation is only $0.20$, versus $0.68$--$0.71$ for regression-based alternatives. This contrast likely reflects the objectives: InfoNCE~\cite{oord2018representation} directly preserves the relative geometry required for identification, whereas correlation and mean-squared error emphasize dimension-wise agreement. Per-subject results, peak relevance, and the full decoder comparison appear in Tables~\ref{tab:persubj} and~\ref{tab:peakrel} and Fig.~\ref{fig:decoder}.
\begin{figure}[!htb]\centering
  \includegraphics[width=0.6\linewidth]{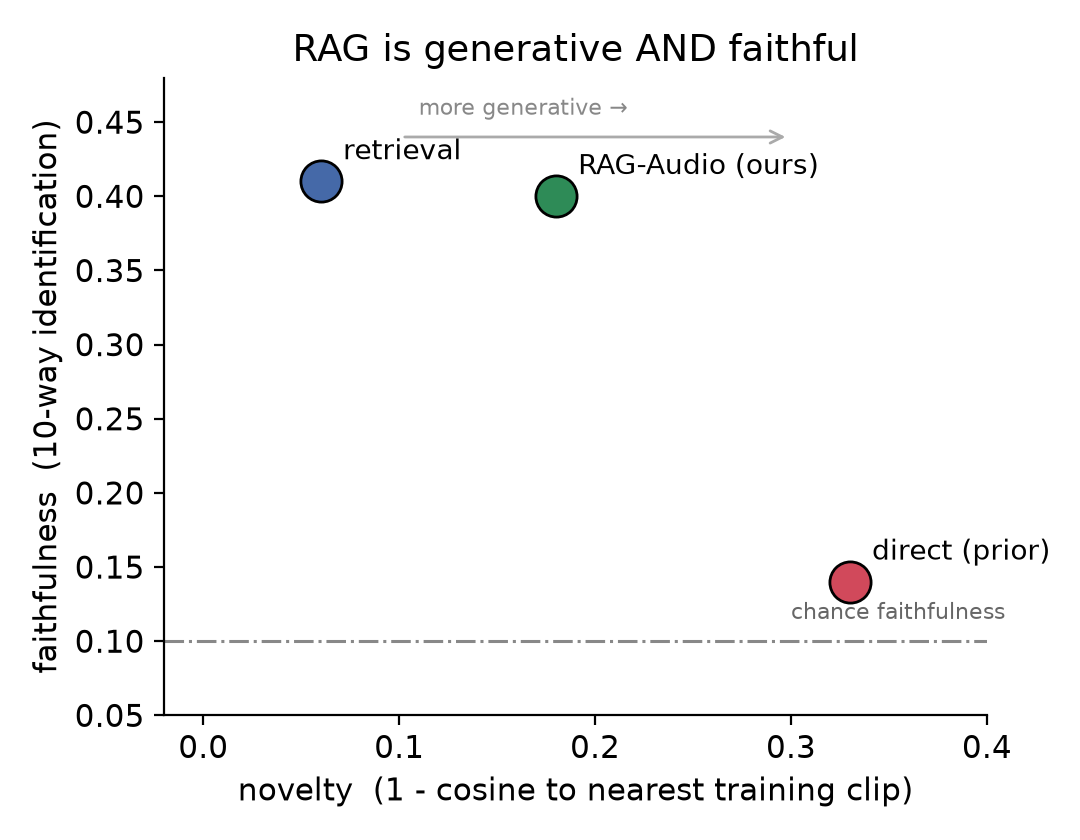}
\caption{Faithfulness (identification) versus novelty: direct is novel but unfaithful, retrieval faithful but verbatim, RAG-Audio faithful and generative.}
  \label{fig:quad}
\end{figure}
The selected voxels also exhibit an anatomically plausible distribution. Figure~\ref{fig:atlas} shows that the relevance scores concentrate in the bilateral superior temporal gyrus, a region associated with auditory processing. Under the Harvard--Oxford atlas~\cite{desikan2006automated}, 78\% of the top-ranked voxels fall within auditory cortex across the two hemispheres. This localization supports the interpretation that the decoder relies primarily on stimulus-related auditory activity rather than scanner artefacts or head motion.
\paragraph{Prior Domination} Direct brain-conditioned generation discards much of the stimulus information recovered by the decoder. Although the decoded embedding reaches 0.43 identification accuracy, generation reduces performance to 0.14 with AudioLDM, 0.14 with TangoFlux, and 0.18 with MusicGen, only modestly above the 0.10 chance level. Table~\ref{tab:main} reports the full comparison, with the complete $N$-way breakdown in Table~\ref{tab:idfull}. The MusicGen result re-implements the decode-then-generate formulation of \cite{denk2023brain2music}.
The degradation is not confined to one generator family. It appears in latent diffusion, rectified flow, and autoregressive generation, while direct-generation FAD ranges from 7.89 to 13.50. Taken together, these results are consistent with prior domination as a property of the decode-then-generate recipe: the decoded condition remains informative, but the pretrained generator produces audio governed largely by its own prior.

 \begin{figure}[!htb]
     \centering
     \includegraphics[width=0.9\linewidth]{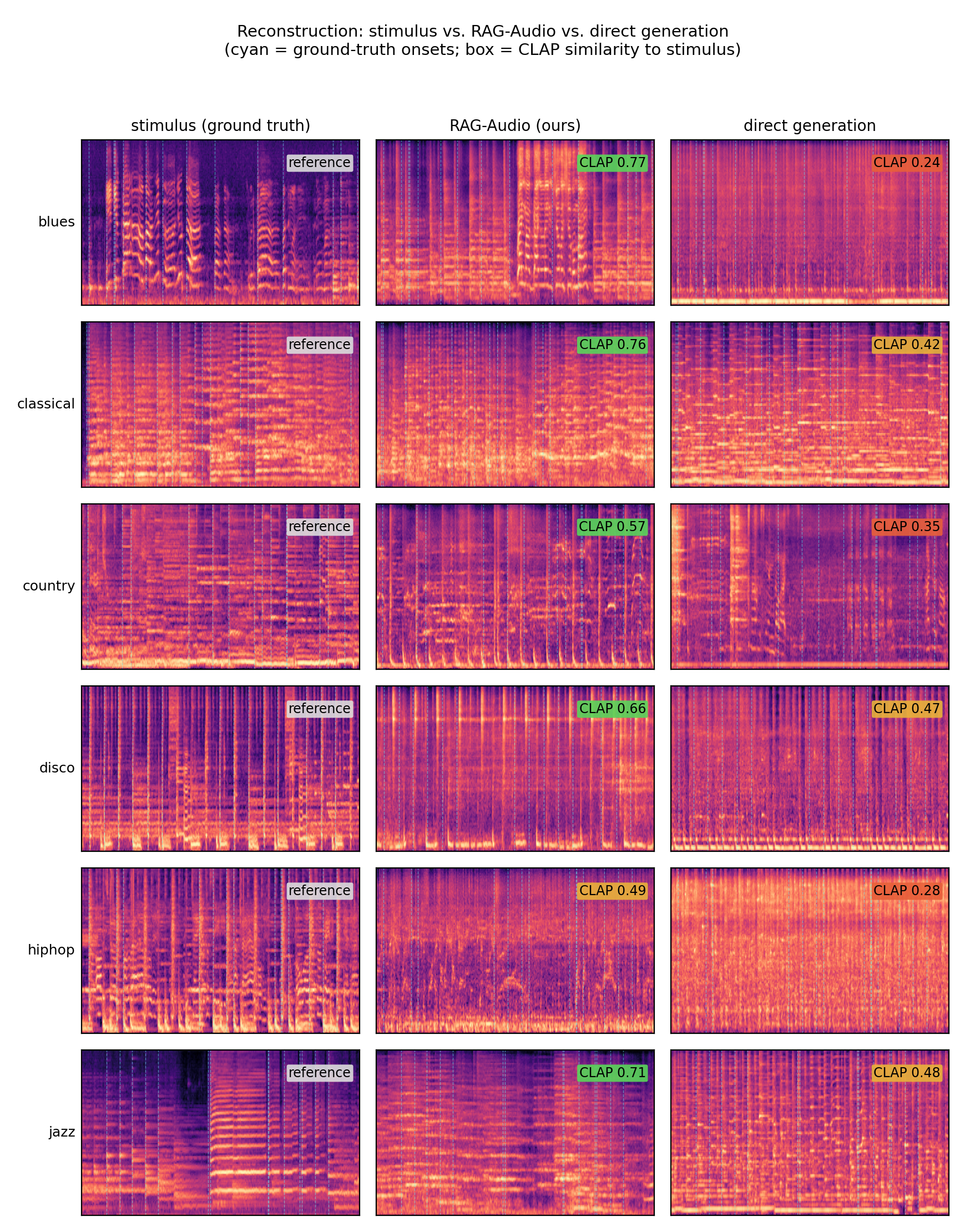}
     \caption{Per-genre reconstructions: stimulus, RAG-Audio, and direct generation (log-mel spectrograms). Cyan lines are stimulus onsets (shared per row); the corner box is CLAP similarity to the stimulus (green/orange/red). Direct generation is realistic but off-grid and low-similarity; RAG-Audio recovers onset and harmonic structure (on-grid, high) without copying any retrieved clip. Structural, not sample-accurate, closeness.}
     \label{fig:recon}
 \end{figure}
\paragraph{Exemplar Anchoring} Exemplar anchoring restores identification to the level of the retrieval reference for latent-diffusion generators. RAG-Audio reaches 0.43 with AudioLDM and 0.40 with TangoFlux, compared with 0.14 for their directly conditioned counterparts and 0.10 chance. These values approximately match both the 0.40 retrieval baseline and the 0.43 decoder-level identification score, corresponding to an approximately $3\times$ improvement over direct generation. The same intervention reduces AudioLDM FAD from 13.50 to 1.25, an approximately order-of-magnitude change, and TangoFlux FAD from 7.89 to 2.36 (Table~\ref{tab:main}); the same reductions hold for FD and KL (Table~\ref{tab:realism}).
This comparison should not be read as RAG-Audio outperforming retrieval. Retrieval directly returns real training audio and is therefore the strongest non-generative reference, with an FAD of 0.89. RAG-Audio does not surpass this value, nor do we claim that it does. Its advantage is instead that it matches retrieval-level identification while producing an edited sample rather than returning the retrieved clip verbatim. Figure~\ref{fig:quad} reflects this distinction: novelty increases from approximately 0.06 for retrieval to 0.18 for RAG-Audio while identification remains at the retrieval level (Table~\ref{tab:novelty}). A genre-level breakdown of the residual errors (Appendix~\ref{app:genre}, Table.~\ref{tab:bank}) shows they remain musically coherent, concentrating among acoustically similar GTZAN genres~\cite{tzanetakis2002musical}.

The anchoring strength controls this balance. As shown in Fig.~\ref{fig:curve} (full sweep in Table~\ref{tab:sweep}), identification decreases smoothly from the retrieval regime toward the 0.10 chance level as the strength approaches 1, while novelty increases as more exemplar structure is removed. The range $s\approx0.25$--$0.40$ provides the most useful operating region in our experiments, retaining stimulus-related structure without reducing generation to direct retrieval. The anchoring advantage is also stable across memory-bank sizes (Appendix~\ref{app:bank}, Table~\ref{tab:bank}).
\paragraph{Mechanism Control} The autoregressive negative control indicates that exemplar availability alone is insufficient. On MusicGen, adding the retrieved exemplar changes identification only from 0.18 to 0.20, in contrast to the approximately $3\times$ gains observed for AudioLDM and TangoFlux in Fig.~\ref{fig:cross}. MusicGen has no continuous latent trajectory that admits intermediate-time initialization; the exemplar can enter only through melody conditioning, which remains governed by the autoregressive token prior.

FAD nevertheless improves modestly, from 8.06 to 6.21. This pattern suggests that the exemplar can nudge acoustic realism without restoring stimulus identity. The cross-generator comparison therefore points to intermediate-time trajectory initialization, rather than the mere presence of a retrieved example, as the mechanism responsible for the faithfulness gains.

\paragraph{Qualitative reconstructions.} Figure~\ref{fig:recon} makes prior domination and its mitigation directly visible. Direct brain-conditioned generation produces spectrograms with plausible spectrotemporal texture, yet their energy drifts off the stimulus onset grid (cyan) and their CLAP similarity to the stimulus stays low across all six examples (red/orange boxes). The failure is thus easy to miss from realism or casual listening alone precisely the regime we call prior domination. Exemplar anchoring recovers the structure the prior discards: RAG-Audio reconstructions align with the onset grid and reproduce the harmonic banding of the stimulus, and the CLAP-similarity boxes rise correspondingly (green), roughly doubling over direct generation across the shown genres. At the same time, the RAG panels differ in fine detail from both the stimulus and any single retrieved clip, consistent with the novelty results in Table~\ref{tab:novelty} and confirming that anchoring edits rather than copies. 
\section{Discussion}
Our results support a simple account of when and why brain-to-audio generation fails. A pretrained generator encodes a strong prior over natural audio, whereas the brain-derived condition is weak and noisy; combined through direct conditioning, the prior dominates and the output drifts toward a fluent but stimulus-agnostic sample. Anchoring the sampling trajectory on a retrieved real-audio exemplar tethers the output to stimulus-related structure, converting
prior domination into a controllable faithfulness-novelty trade-off governed by the anchoring strength.

This mitigation is deliberately generator-specific, and the specificity is informative rather than incidental. Because exemplar anchoring acts by initializing a continuous latent trajectory, it applies to latent-diffusion and rectified-flow generators but not to autoregressive models, which expose no such trajectory. The MusicGen negative control makes this concrete: supplying the identical exemplar raises $10$-way identification only marginally, from $0.18$ to
$0.20$, whereas the same exemplar restores AudioLDM and TangoFlux to retrieval-level accuracy ($0.43$ and $0.40$). Trajectory initialization, not the mere availability of a retrieved clip, is therefore the operative mechanism.
 
We are precise about what RAG-Audio does and does not achieve. It does not surpass plain retrieval on FAD, and it cannot: retrieval returns real training audio, which is FAD-optimal by construction ($0.89$). RAG-Audio's contribution is instead to match retrieval-level identification while emitting a newly generated sample rather than replaying a stored clip, cutting FAD by roughly an order of magnitude relative to direct generation (e.g., AudioLDM $13.5\!\rightarrow\!1.25$). Its niche is exactly the setting that retrieval cannot serve: where novel audio is required and a verbatim training clip is unacceptable, yet faithfulness to the heard stimulus must still be preserved.

\section{Limitations}
Several limitations bound our claims. The study covers a single dataset and a single modality: music decoded from fMRI so transfer to speech or environmental sound remains untested. The memory bank is bounded by the training stimuli; RAG-Audio anchors to, and therefore cannot reconstruct content outside, that support. Exemplar anchoring requires an initializable latent trajectory and thus does not extend to autoregressive generators. Our $N$-way identification measures
semantic agreement in CLAP space rather than sample-accurate waveform recovery, so the spectrograms should not be read as evidence of exact reconstruction. Finally,
the anchoring strength $s$ is a hyperparameter: we report a working range ($s\!\approx\!0.25$--$0.40$) rather than a single optimum, and the best operating
point may differ across generators and datasets.
\section{Conclusion}
We identified \emph{prior domination} as a general failure mode of the decode-then-generate recipe for brain-to-audio reconstruction: a decoded embedding that is itself identifiable ($0.43$ at $10$-way, against $0.10$ chance) can nonetheless collapse to near-chance reconstructions ($0.14$--$0.18$) once a strong generator is applied. We introduced exemplar anchoring, initializing the generator's sampling trajectory from a retrieved real-audio clip, via SDEdit-style noising for diffusion models and the analogous rectified-flow interpolation for flow models and showed that it restores retrieval-level identification while keeping generation genuinely generative and reducing FAD by roughly an order of magnitude. An autoregressive negative control localizes the effect to latent-trajectory initialization rather than retrieval alone. Beyond the method, our single-harness comparison turns a failure mode previously reported only qualitatively in brain-to-image work into a quantitative, generator-general characterization one likely to sharpen as pretrained generators continue to improve.
\section*{Acknowledgements}
This work was supported by the European Union's Horizon Europe research and
innovation programme under the Marie Sk\l{}odowska-Curie grant agreement
No.~101205348 (CASPER). We acknowledge the EuroHPC Joint Undertaking for
awarding this project access to the EuroHPC supercomputer LEONARDO, hosted by
CINECA (Italy) and the LEONARDO consortium, through the EuroHPC AI Factories
"AI for Science and Collaborative EU Projects" Access call (proposal
No.~EHPC-AIF-2026SC01-041). We further acknowledge the CINECA award under the
ISCRA initiative (Class C project IsCd5\_CASPER-A), for the availability of
high performance computing resources and support. Views and opinions expressed
are however those of the author(s) only and do not necessarily reflect those of
the European Union or the European Research Executive Agency. Neither the
European Union nor the granting authority can be held responsible for them.

\bibliographystyle{plainnat}
\bibliography{references}

\clearpage
\appendix
\section{Code and Reproducibility}
\label{app:code}
For reproducibility, we release the full codebase, including preprocessing, the fMRI-to-CLAP decoder, retrieval, exemplar anchoring, and evaluation, in a public repository:\\
\mbox{\url{TBA}}\\The repository documents the environment, checkpoints, and configuration files needed to reproduce every table and figure in this document.
\section{Use of Large Language Models}
\label{app:llm}
Large language models were used only to improve grammar and clarity in author-written text.
\section{Implementation Details}
\label{app:impl}

\paragraph{Decoder.} The contrastive decoder $f_\theta$ is a multilayer perceptron mapping the $2000$ selected voxels to the $512$-dimensional CLAP space. It is trained for $400$ epochs with the InfoNCE objective~\cite{oord2018representation} at temperature $\tau=0.07$, using the Adam optimizer with weight decay; full layer widths, learning rate, and batch size are given in our released configuration. Voxels are selected by the relevance score of Eq.~\ref{eq:vox} ($K=2000$), with a haemodynamic lag of $\ell=3$ TRs and a temporal window of $w=8$ TRs. Motion correction is rigid-body, performed in native space with ANTsPy~\cite{avants2011reproducible}.
\paragraph{Generators.} All three generators are frozen and used at their public checkpoints: AudioLDM (\texttt{cvssp/audioldm-s-full v2})~\cite{liu2023audioldm}, TangoFlux (\texttt{declare-lab/TangoFlux})~\cite{hung2024tangoflux}, and MusicGen (\texttt{facebook/musicgen-melody})~\cite{copet2023simple}. For anchored generation we sweep the strength grid $s\in\{0.15,0.25,0.35,0.5,0.65,0.8,1.0\}$; generation step counts and guidance scales are listed in the released configuration. Identification uses the LAION CLAP checkpoint \texttt{laion/clap-htsat-unfused}~\cite{wu2023large}, which is distinct from the CLAP variant used as the decoder target.
\section{Decoder Baselines}
\label{app:decoder}
\begin{figure}[!htb]\centering
  \includegraphics[width=0.5\linewidth]{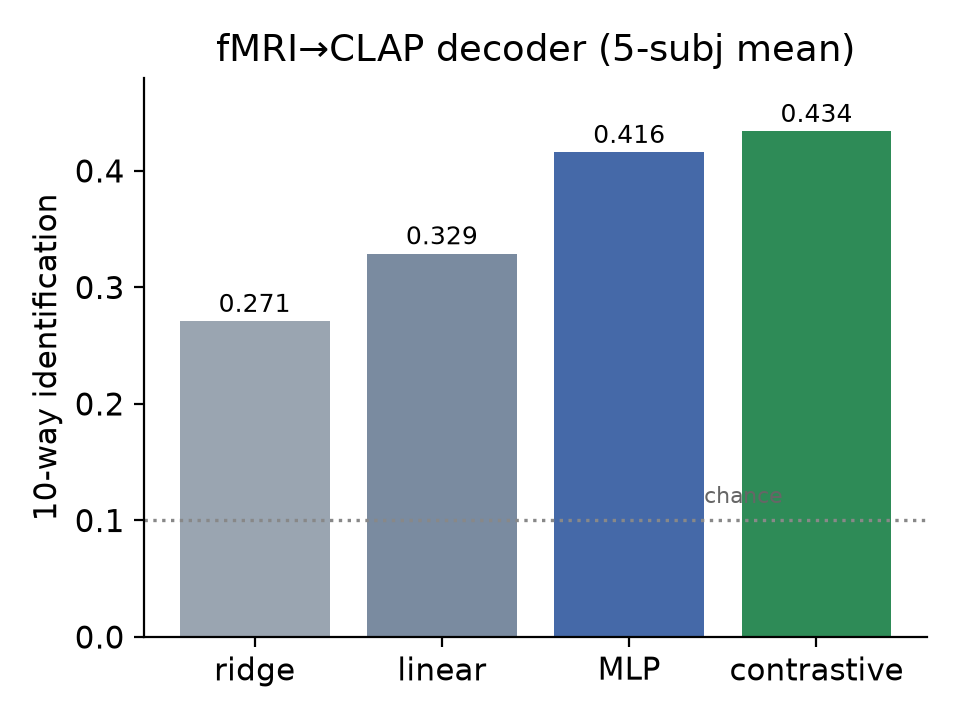}
  \caption{Decoder 10-way identification (5-subj mean).}
  \label{fig:decoder}
\end{figure}

The contrastive decoder is compared against ridge, linear, and MLP regressors trained to predict the CLAP target directly. As summarized in Figure~\ref{fig:decoder}, the contrastive decoder attains the best $10$-way identification despite the \emph{lowest} dimension-wise correlation with the target embedding (Pearson $r=0.20$, versus $0.68$--$0.71$
for the regression decoders). This apparent paradox reflects the training objective: InfoNCE optimizes the relative cosine geometry that $N$-way identification actually measures, whereas ridge/linear/MLP minimize dimension-wise error, which need not preserve nearest-neighbor rankings. Because retrieval and identification both operate on cosine similarity, the contrastive criterion is the better match to the downstream task.
\begin{table}[!htb]
\centering
\small
\caption{Identification at $N\!=\!2,5,10$ for every generator $\times$ arm (direct, RAG, retrieval; $n{=}300$, chance $1/N$). RAG matches the retrieval bound for the latent-trajectory generators but not MusicGen. $\dagger$: only $10$-way was
aggregated for AudioLDM (Table~\ref{tab:main}).}
\label{tab:idfull}
\begin{tabular}{llccc}
\toprule
Generator & Arm & 2-way & 5-way & 10-way\\
\midrule
\multirow{3}{*}{AudioLDM}
 & direct    & $\dagger$ & $\dagger$ & 0.14\\
 & RAG       & $\dagger$ & $\dagger$ & 0.43\\
 & retrieval & $\dagger$ & $\dagger$ & 0.40\\
\multirow{3}{*}{TangoFlux}
 & direct    & 0.515 & 0.249 & 0.140\\
 & RAG       & 0.785 & 0.555 & 0.403\\
 & retrieval & 0.792 & 0.554 & 0.406\\
\multirow{3}{*}{MusicGen}
 & direct    & 0.590 & 0.310 & 0.180\\
 & RAG       & 0.615 & 0.336 & 0.201\\
 & retrieval & 0.796 & 0.553 & 0.402\\
\bottomrule
\end{tabular}
\end{table}
\FloatBarrier
\section{Per-Subject Results}
\label{app:persubj}
 The decoder is reliable across all five subjects. Table~\ref{tab:persubj} reports per-subject identification for the contrastive decoder ($60$ test clips each): every subject exceeds chance at all $N$, with a mean $10$-way accuracy of $0.434$ (chance $0.10$) and a best subject (sub-003) at $0.532$. Table~\ref{tab:peakrel} reports the peak voxel relevance ($\max|\mathrm{corr}|$ with the CLAP target) per subject, ranging from $0.425$ (sub-001) to $0.599$ (sub-003); the ordering tracks identification accuracy, indicating that subjects with stronger auditory-cortex coupling are decoded more reliably. Bilateral superior-temporal-gyrus localization is consistent across subjects (Figure~\ref{fig:atlas}), and the single-subject map for sub-001 is shown in the main text.
\begin{table}[!htb]\centering\small
\caption{Per-subject identification of the contrastive decoder ($60$ test clips each; chance $1/N$). All five subjects exceed chance at every $N$, confirming the decoder is reliable across subjects.}
\label{tab:persubj}
\begin{tabular}{lccccc}
\toprule
Subject & 2-way & 5-way & 10-way & 50-way & Pearson\\
\midrule
sub-001 & 0.782 & 0.555 & 0.393 & 0.121 & 0.177\\
sub-002 & 0.820 & 0.591 & 0.421 & 0.145 & 0.189\\
sub-003 & 0.836 & 0.651 & 0.532 & 0.206 & 0.234\\
sub-004 & 0.790 & 0.560 & 0.437 & 0.173 & 0.174\\
sub-005 & 0.794 & 0.546 & 0.388 & 0.154 & 0.233\\
\midrule
mean    & 0.804 & 0.581 & 0.434 & 0.159 & 0.201\\
\bottomrule
\end{tabular}
\end{table}
\begin{table}[!htb]\centering\small
\caption{Per-subject peak voxel relevance (max $|\mathrm{corr}|$ with the CLAP target), indexing decoder strength in auditory cortex for each subject.}
\label{tab:peakrel}
\begin{tabular}{lccccc}
\toprule
Subject & sub-001 & sub-002 & sub-003 & sub-004 & sub-005\\
\midrule
peak $|$corr$|$ & 0.425 & 0.496 & 0.599 & 0.454 & 0.485\\
\bottomrule
\end{tabular}
\end{table}
\begin{table}[!htb]
\centering
\small
\caption{Audio realism ($n{=}300$; lower is better): FAD (VGGish) with FD and KL (PANNs Cnn14). RAG reduces all three for AudioLDM and TangoFlux toward the real-audio retrieval reference, while MusicGen changes little.}
\label{tab:realism}
\begin{tabular}{llccc}
\toprule
Generator & Arm & FAD & FD & KL\\
\midrule
\multirow{3}{*}{AudioLDM}
 & direct    & 13.49 & 79.40 & 3.08\\
 & RAG       & 1.25  & 17.23 & 1.07\\
 & retrieval & 0.89  & 16.40 & 1.05\\
\multirow{3}{*}{TangoFlux}
 & direct    & 7.89 & 59.11 & 2.24\\
 & RAG       & 2.36 & 21.49 & 1.01\\
 & retrieval & 0.89 & 16.40 & 1.05\\
\multirow{3}{*}{MusicGen}
 & direct    & 8.06 & 52.99 & 1.85\\
 & RAG       & 6.21 & 59.32 & 1.82\\
 & retrieval & 0.89 & 17.55 & 1.08\\
\bottomrule
\end{tabular}
\end{table}
\begin{table}[!htb]
\centering
\caption{Novelty ($1-\cos$ to the nearest training clip) and pairwise diversity for the TangoFlux arms. RAG is far more novel than verbatim retrieval (${\approx}0$) while staying below unconstrained direct generation.}
\label{tab:novelty}
\begin{tabular}{lcc}
\toprule
Arm & novelty & diversity\\
\midrule
retrieval        & 0.06 & 0.42\\
RAG (ours)       & 0.18 & 0.39\\
direct (prior)   & 0.33 & 0.48\\
\bottomrule
\end{tabular}
\end{table}
\FloatBarrier
\section{Memory-Bank Size Ablation}
\label{app:bank}
\section{Genre Confusion}
\label{app:genre}
\begin{figure}[!htb]
\centering
  \includegraphics[width=0.7\linewidth]{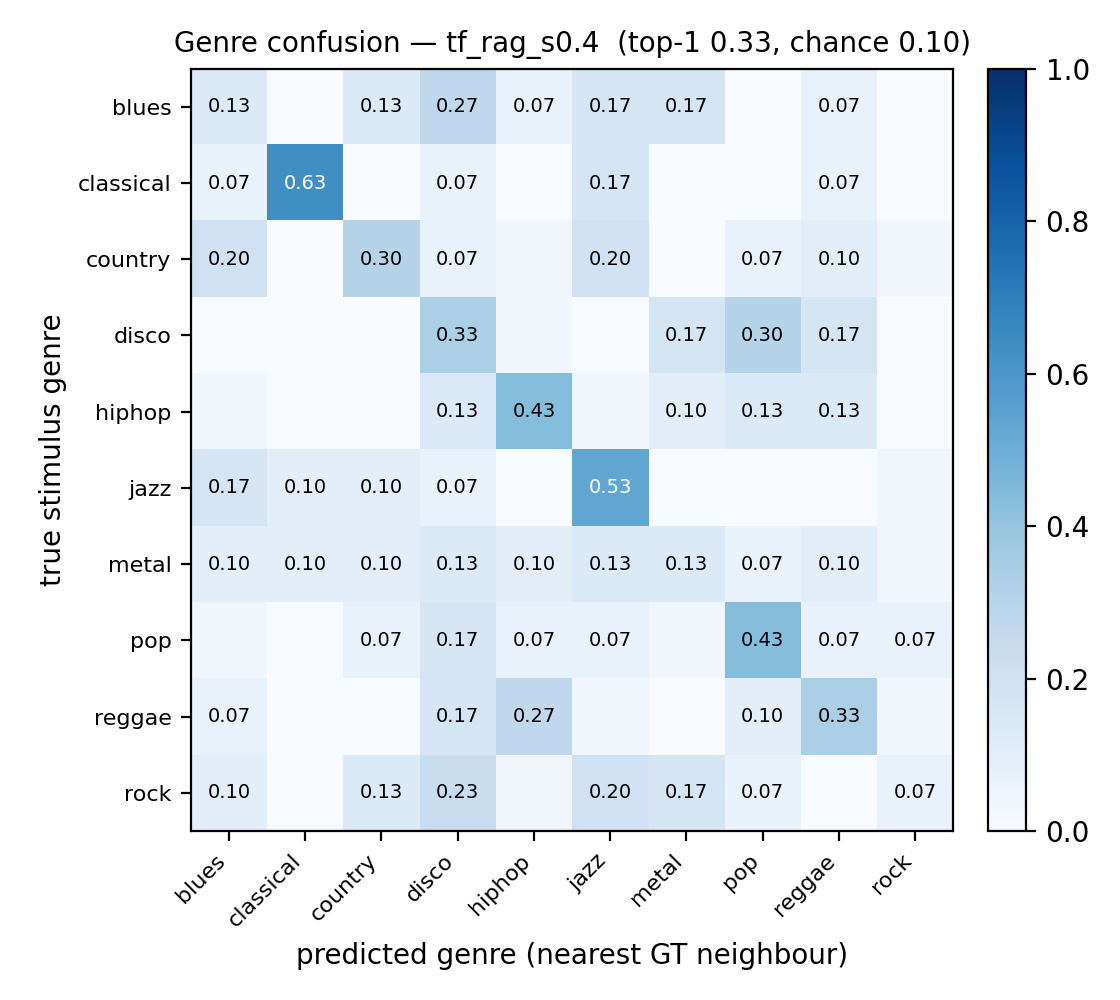}
  \caption{Genre confusion for the RAG arm (TangoFlux).}
  \label{fig:genre}
\end{figure}

To characterize the residual errors, each generated clip is labelled by the genre of its nearest ground-truth neighbour in CLAP space (leave-one-out). Figure~\ref{fig:genre} shows the resulting confusion matrix for the RAG arm (TangoFlux): top-1 genre accuracy is $0.33$, or $3.3\times$ the $0.10$ chance level. Well-separated genres are recovered most often (classical $0.63$, jazz $0.53$, hip hop/pop $0.43$), while the errors concentrate among the acoustically overlapping GTZAN~\cite{tzanetakis2002musical} genres (metal, rock, blues) that are confusable even for audio-only classifiers. The confusions are musically coherent (e.g., disco$\leftrightarrow$pop$\leftrightarrow$reggae, rock$\leftrightarrow$disco), indicating that reconstruction errors preserve broad timbral and rhythmic structure rather than being random.
\FloatBarrier
\section{SDEdit Strength Sweep}
\label{app:sweep}
Table~\ref{tab:sweep} details the dependence of $10$-way identification on the anchoring strength $s$ that underlies Figure~\ref{fig:curve}. For AudioLDM, identification is highest at low strength ($0.49$ at $s=0.2$) and decreases
monotonically toward the $0.10$ chance level as $s\to1$, crossing the retrieval bound within the working range. TangoFlux was evaluated at its operating point $s=0.4$, where it reaches $0.40$. Realism is stable across the anchored range: AudioLDM FAD is $1.14$, $1.12$, and $1.25$ at $s=0.15$, $0.2$, and $0.25$ respectively, so faithfulness can be tuned over this interval without degrading audio quality. We therefore adopt $s\approx0.25$--$0.40$ as the default operating region.
\begin{table}[!htb]\centering\small
\caption{Ten-way identification versus anchoring strength $s$ (AudioLDM sweep; TangoFlux at its operating point). Accuracy is highest at low $s$ and decays to the
$0.10$ chance level as $s\!\to\!1$; \emph{retr.}\ = retrieval bound, \emph{direct} = prior-only.}
\label{tab:sweep}
\begin{tabular}{lccccccc}
\toprule
Generator & $s{=}0.2$ & 0.25 & 0.3 & 0.4 & 0.5 & retr. & direct\\
\midrule
AudioLDM  & \textbf{0.49} & 0.43 & 0.39 & 0.30 & 0.14 & 0.40 & 0.14\\
TangoFlux & --- & --- & --- & 0.40 & --- & 0.40 & 0.14\\
\bottomrule
\end{tabular}
\end{table}

We tested whether RAG-Audio's advantage grows as the retrieval memory shrinks, hypothesizing graceful degradation. The data did not support this. Table~\ref{tab:bank} report AudioLDM RAG ($s=0.2$) against retrieval FAD at bank sizes $30$, $120$, and $480$. At reduced bank sizes the estimates are noisy ($n=60$ pairs versus $300$ at the full bank), and RAG
tracks retrieval within that noise rather than separating from it as the bank shrinks. We report this as an honest negative result: the benefit of anchoring does not increase under retrieval scarcity in this dataset.
\begin{table}[!htb]
\centering
\small
\caption{Memory-bank size ablation: AudioLDM RAG ($s{=}0.2$) vs.\ retrieval FAD at bank sizes $30/120/480$ ($n{=}60$ pairs, noisy). RAG tracks retrieval within noise and does not degrade gracefully as the bank shrinks.}
\label{tab:bank}
\begin{tabular}{lccc}
\toprule
 & bank 30 & bank 120 & full (480)\\
\midrule
RAG FAD       & 1.43 & 1.88 & 1.12\\
retrieval FAD & 1.90 & 1.28 & 0.89\\
\bottomrule
\end{tabular}
\end{table}
\FloatBarrier
\section{Dataset and Reproducibility}
\label{app:repro}
We use the Brain2Music fMRI dataset~\cite{denk2023brain2music}, released as OpenNeuro \texttt{ds003720}~\cite{nakai2022music}, obtained via DataLad (\texttt{datalad get} on the \texttt{*\_bold.nii} files); the auditory stimuli are drawn from the ten GTZAN genres~\cite{tzanetakis2002musical}. We follow the official subject-wise split of $480$ training and $60$ test clips per subject.
Preprocessing applies ANTsPy rigid motion correction in native space~\cite{avants2011reproducible}, a lag of $3$ TRs and window of $8$ TRs, and top-$2000$ voxel selection by Eq.~\ref{eq:vox}. Auditory-cortex overlap is computed against the Harvard-Oxford atlas~\cite{desikan2006automated}. All runs use a single GPU per job; we release seeds, code, and configurations.

\end{document}